\documentclass 
{aa}

\usepackage{aabib99}

\usepackage{graphicx}
\newcommand{\apjl}{ApJ Let.}
\newcommand{\apj}{ApJ}

\newcommand{\aap}{A\&A}

\newcommand{\be}{\begin{equation}}
\newcommand{\ee}{\end{equation}}

\begin{document}

\title{Rotating compact strange stars}

\titlerunning{Rotating compact strange stars}

\author{Dorota Gondek-Rosi{\'n}ska$^1$, Tomasz Bulik$^1$, Leszek
Zdunik$^1$, Eric Gourgoulhon $^2$, Subharthi Ray $^{3,4}$, Jishnu Dey  $^3$ and
Mira Dey$^4$ }

\authorrunning{Gondek-Rosi{\'n}ska et~al.}
\institute{
 $^1$Nicolaus Copernicus Astronomical Center,
Bartycka 18, 00-716 Warszawa,Poland\\
$^2$D{\'e}partement d'Astrophysique Relativiste et de Cosmologie 
 UPR 176 du CNRS, Observatoire de Paris, F-92195 Meudon Cedex, France\\
$^3$ Abdus Salam ICTP,
Trieste, Italy and Azad Physics Centre, Maulana Azad College, Calcutta
700013, India, azad@vsnl.com \\
$^4$ Dept. of Physics, Presidency College,
Calcutta 700 073, India and Abdus Salam ICTP, Trieste, Italy, \\
\\ permanent address : 1/10 Prince Golam Md. Road, Calcutta 700
026,India}


\date{Received ................, Accepted ....................}

\thesaurus{??????????????}
 
\maketitle 
 
\begin{abstract} 
  { We compute numerical models of uniformly rotating strange stars in
    general relativity for the recently proposed QCD-based equation of
    state (EOS) of strange quark matter \cite{1998PhLB..438..123D}.
    Static models based on this EOS are characterised by a larger
    surface redshift than strange stars within the MIT bag model. The
    frequencies of the fastest rotating configurations described by
    Dey model are much higher than these for neutron stars models and
    for the simplest strange star MIT bag model. We determine a number
    of physical parameters for such stars and compare them with those
    obtained for neutron stars.  We construct constant baryon mass
    equilibrium sequences both normal and supramassive.  Similarly to
    the neutron star model a supramassive strange star, prior to
    collapse to a black hole, spins up as it loses angular momentum.
    We find the upper limits on maximal masses and maximal frequencies
    of the rotating configurations.  There are two regimes for maximal
    mass of rotating configurations as the rotation frequency
    increases; first, for low rotation frequencies the increase in the
    maximal mass is small ($M_{\rm max}(f)$ is very close to the line
    of limiting stability against quasi-radial perturbations limit);
    second, for higher frequencies the maximal mass configurations are
    Keplerian ones. We show that the maximal rotating frequency for
    all considered baryon mass is never the Keplerian one.  A normal
    and low mass supramassive strange stars gaining angular momentum
    always slows down just before reaching the Keplerian limit. For a
    high mass supramassive strange star sequence the Keplerian
    configuration is the one with the lowest rotational frequency in
    the sequence.  The value of $T/W$ for rapidly rotating strange
    stars of any mass is significantly higher than those for ordinary
    neutron stars. For Keplerian configurations it increases as mass
    decreases. The results are robust for all linear self-bound
    equations of state.  }

\end{abstract}

\keywords{dense matter - equation of state - stars: neutron stars: 
pulsars: general}

\section{Introduction}

The idea of existence of quark matter dates back to the early
seventies. \cite{1971Bodmer} remarked that stable matter
consisting of deconfined up, down and strange quarks may be
stable.  If that was the case then object made of such matter -
strange stars could exist \cite{1984Witten}.  Most of the
previous calculations of strange stars properties (static models
- see e.g. \cite{1986ApJ...310..261A,1986A&A...160..121H},  for rotating
models - see references in  Gourgoulhon et al. 1999) were done
for an equation of state based on the MIT bag model, in which
quark confinement and asymptotic freedom were postulated from
the very beginning, and  the deconfinement of quarks at high
densities was not obvious. 

Quite a number of authors have discussed the implications of existence
of strange stars. Alpar (1987) \nocite{1987Alpar} noted that 
glitching radio pulsars
are not strange stars.  Since a binary merger with a strange star
might contaminate the entire Galaxy with strange matter seeds
\cite{1988PhRvL..61.2909M,1991CaldFried} concluded that strange stars
can not be formed directly in supernovae. Strange stars could then
exist as millisecond pulsars and be formed from neutron stars through
a phase transition \cite{1994AiA...286L..17K,1996PhRvL..77.1210C}.

Strange stars described by the simple MIT bag with massless and
noninteracting  quarks have orbital frequencies in the marginally
stable orbit which are higher than the lowest maximum frequency
observed of kHz quasiperiodic oscillations observed in LMXBs
\cite{1999A&A...344L..71B}. However the
corresponding frequencies for more sophisticated models (MIT bag
model with massive strange quarks and lowest order QCD
interactions, and/or rotation (Stergioulas et al., 1999, Zdunik et al., 2000)
 do allow frequencies as low as 1 kHz, in
agreement with observations.

Recently Dey et al. (1998) derived an EOS for strange
matter which has asymptotic freedom built in and  describes
deconfined quarks at high density and confinement at zero
density.  This model, with an appropriate choice of the EOS
parameters, gives absolutely stable strange quark matter.  This
equation of state was used to calculate the structure of static
strange stars and  the mass-radius relations. Later,  it was
suggested \cite{1999PhRvL..83.3776L} that the millisecond X-ray
pulsar SAX1808.4-3658 is a strange star. Also this equation
allowed to explain the observed properties of other objects: an
analysis of semi-empirical mass-radius relations in Her~X-1, 4U
1728-34 \cite{1999ApJ...527L..51L} and 4U~1820-30 leads to the
suggestion that these objects host strange stars
\cite{1998PhLB..438..123D}.  Two cases of this model have been
used in these papers, which will be denoted as SS1 and SS2
equations of state.  They both give a rather low value of the maximum
gravitational mass $M_{max}=1.33\,M_\odot$ for SS1 and $
M_{max}=1.44\,M_\odot $ for SS2.  Very recently the model of 
\cite{1998PhLB..438..123D} was used for calculating frequencies of
marginally stable orbits around static strange stars and strange
stars rotating with frequency 200 and 580 Hz (Datta et
al. 2000). The authors conclude that very high QPO
frequencies in the range of $1.9$-$3.1$\, kHz would imply existence of a
non-magnetized strange star X-ray binary rather than a neutron
star X-ray binary.  

Up to present no systematic study of rotating strange stars
given by the model of \cite{1998PhLB..438..123D} has been done. 
In this  paper we construct constant baryon mass equilibrium
sequences both normal and supramassive.  Calculations of
equilibrium sequences of rapidly rotating compact stars are
crucial to understand various astrophysical phenomena and
objects, such as LMXBs, QPO and millisecond pulsars. In this paper
we calculate the properties of rapidly rotating strange stars
described by the models SS1 and SS2. We study which of them are
characteristic for stars within the Dey et al. 1998 model and
which are common for all models described by self-bound EOS. We
compare the properties of compact rotating strange stars with those
for neutron stars.  We find the upper limits on observable
astrophysical quantities, such as masses and frequencies of
rotating stars. We locate the rotating maximum mass model and
the maximum angular velocity model.

In section 2 we outline briefly the equation of state used
throughout this work and compare static model SS1 and SS2 stars
with the MIT bag model strange stars. In section 3 we describe
the rotating configuration of the compact strange stars, and in
section 4 we discuss the results.

\section{Equation of state and static strange star models}

In present paper we describe the strange quark matter by the
model presented by Dey et al. (1998). In this model
quarks of the density dependent mass are confined at zero
pressure and deconfined at high density. The quark interaction
is described by an interquark vector potential originating from
gluon exchange, and by a density dependent scalar potential
which restores the chiral symmetry at high densities.

To calculate  the rotating stellar configurations 
 we use  linear
approximation of the equations of state. We linearize the
dependence $P(\rho)$ by the function:  
\begin{equation}
p=a\cdot (\rho-\rho_0).
\label{linea}
\end{equation}
with $n(P)$ resulting from first law of thermodynamics (Zdunik, 2000).
In general the equation of the type (\ref{linea}) corresponds to a
self-bound matter at the density (mass-energy) $\rho_0$ at zero
pressure and with a fixed sound velocity ($\sqrt{a}$).  All stellar
parameters are subject to well known scaling relation with appropriate
powers of $\rho_0$ for a fixed value of  $a$ (see e.g. Witten 1984; 
Zdunik, 2000) 

To calculate parameters $a$ and $\rho_0$ we use least squares fit
method taking into account the region of the densities which is
relevant to the interior of stable stellar configurations, i.e we
neglect in the fitting the part of the EOS for densities larger than
the central density in the last stable configuration (maximum mass in
the non rotating case).  For the case of SS1 we obtained the values of
$a=0.463$, $\rho_0=1.15\times 10^{15}~{\rm g\,cm^{-3}}$,
$n(P=0)=n_0=0.725~{\rm fm}^{-3}$ and for the case SS2 the values are
$a=0.455$, $\rho_0=1.33\times 10^{15}~{\rm g\,cm^{-3}}$,
$n_0=0.805~{\rm fm}^{-3}$.

We have calculated the static star configurations using both the
tabulated  equation of state of \cite{1998PhLB..438..123D} and
its linear approximation. The linear approximation agrees very
well  with calculations of stellar parameters of non rotating
configurations using the tabulated form of the  EOS. The
difference in the whole range of masses and radii is smaller
than 2\%, and the maximum mass point agrees  within  0.2\%.

We present the physical parameters for the maximum mass static strange
stars described by equations SS1, SS2 in the left columns of Table 1.
The stars described by EOS SS1 and SS2 are very compact i.e.  the
gravitational redshifts $z$ for the maximum mass configurations are
much larger than those for SS within the MIT Bag model (also larger
than $z$ for most models of neutron stars), for which $z$ varies from
0.432 to 0.477 for massive strange quark with $m_s=250\ $MeV and for
massless quarks respectively. The maximal baryon mass in the case of
static strange stars described by the SS1 and SS2 is relatively low
(most of the neutron star models have maximum baryon mass close to $2\ 
M_\odot$. In the case considered here the difference between the
baryon mass and the gravitational mass is much higher ($24 \%$ and
$29\% $) than in the case of neutron stars (up to $10\ \%$) (for the
MIT bag stars it is $\sim \ 20-34\ \% $). It should be however noted
that in the case of strange stars we calculate total baryon mass of
the star using nucleon mass and thus we include the binding energy of
strange matter with respect to nuclear one.
 
\begin{table}
\begin{tabular}{lrrrr}
  & \multicolumn{2}{c}{\rm static models} & 
\multicolumn{2}{c}{\rm rotating models} \\ \hline 
Model:  & SS1 & SS2 & SS1 & SS2   \\ \hline \hline \\
$M[M_\odot]$  &1.435$ $ & $1.323 $ & 2.04 & 1.88 \\
$M_{\rm bar}[M_\odot]$  & $1.853$  & $1.641$  & 2.60  & 2.29 \\ 
$R_{eq}$[km]          & $7.07$ & $6.55$ & 10.5 & 10.0\\ 
$n_c$[fm$^{-3}$] & $2.35$ & $2.638$ &  1.75 & 1.77 \\ 
$\rho_c$[$10^{15}$ g\,cm$^{-3}$]
               & $4.68$  & $5.5$ & 3.245   & 3.219 \\
$P[{\rm ms}]$     &  - & - & 0.395  & 0.381 \\ 
$z_{eq}^f$            & $0.5798$ & $0.5741$ & - 0.382 & - 0.376 \\
$z_{eq}^b$            & $0.5798$ & $0.5741$ & 3.633 & 3.284 \\
$z_{pole}$            & $0.5798$ & $0.5741$ & 1.038 & 0.963 \\     
\end{tabular} 
     
{\caption {\small Properties of the strange stars within Dey model with 
maximal masses. The symbols are as follows:
$M$ and $M_{\rm bar}$ are gravitational and baryon masses respectively, 
$R_{eq}$ is circumferential radius; $n_{\rm c}$ is 
the central baryon density; $\rho_{\rm c}$ is the central proper energy 
density devided by $c^2$, $P$ is the rotation period; $z_{eq}^f$, $z_{eq}^b$,
$z_{pole}$ are the redshift for an emission at the equator and in the 
direction of rotation, the redshift for an emission at the equator and in 
the direction opposite to rotation and the redshift at the stellar pole 
respectively.}
}
\label{tab1}
\end{table}

\section{Rotating star configurations}

We have calculated numerical models of the uniformly rotating
strange stars described by the SS1 and SS2 equations of state
using the multi-domain spectral methods developed by
Bonazzola et al. 1998. This method has been used
previously for calculating rapidly rotating strange stars
described by the MIT Bag model (Gourgoulhon et al. 1999,
Zdunik et al., 2000).  The  multi-domain technique allows to treat
the density discontinuity at the surface of "self-bound"
stars  even with a very high $\rho_0$.

 We
construct equilibrium sequences of rotating compact strange stars
with constant baryon mass, i.e. the   so called evolutionary
sequences (for example a pulsar keeps its baryon mass constant
while slowing down; neglecting accretion a compact star keeps
its rest mass constant during the evolution).  Similarly to
neutron stars, we identify the  normal  and the supramassive
stars.   An  sequence is called normal 
if it   terminates at the zero angular momentum 
 limit  with a  static, spherically symmetric solution,
and it is called a supramassive sequence if it does not.  The boundary
between these two sequences is the sequence with the  maximum
baryon mass of a static configuration. The angular momentum (and
the central density of a star) changes monotonically along each
sequence. Note, that the rotational frequency $f$ does not
necessarily  change monotonically along a sequence, since a star
changes its shape and, consequently,  moment of inertia  with increasing
angular momentum.

\subsection{Equilibrium sequences}

\begin{figure}
\includegraphics[width=0.9\columnwidth]{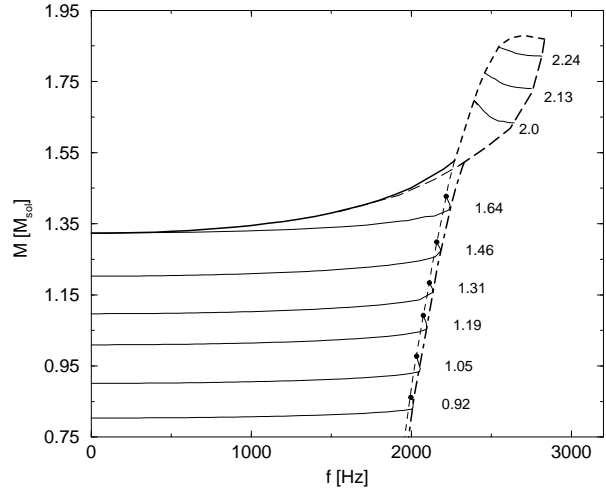}
\caption{Gravitational mass as a function of the rotation frequency for
  SS2 model.  Thin solid lines correspond to evolutionary sequences
  with fixed baryon mass labelled close to each line. The angular momentum 
  increases along each
  curve from $J=0$ for static configurations ($J_ {\rm min}$ for
  supramassive stars) to $J_{\rm max}$ for the Keplerian ones
  represented by filled circles for normal and low massive
  supramassive stars. The thick lines correspond to the upper limits
  on gravitational mass and rotational frequency; the dashed line
  corresponds to the mass-shed limit, the long-dashed line is the
  quasi-radial stability limit, while the dashed-dotted line shows the
  fastest rotating configurations }
\label{Mfdey2}
\end{figure}

\begin{figure}
\includegraphics[width=0.9\columnwidth]{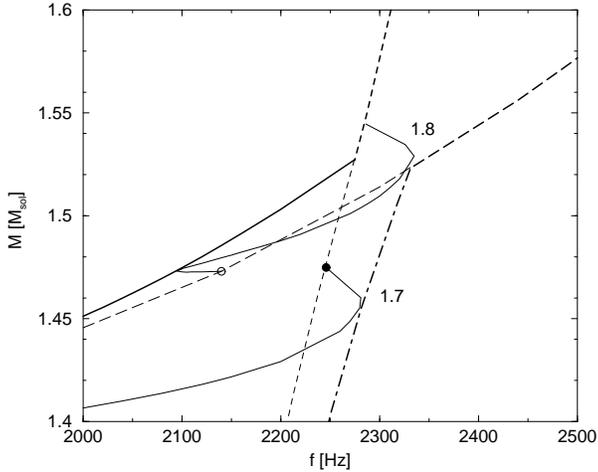}
\caption{Limits on gravitational mass and on the rotation frequency 
  and evolutionary supramassive sequences with rest mass 1.7 and 
  1.8 $M_\odot$.
  All lines are as indicated in Figure ~\ref{Mfdey2}. The marginally
  stable configuration with respect quasi-radial oscillation is shown
  as open circle.  }
\label{Mfdey2cd}
\end{figure}

\begin{figure}
\includegraphics[width=0.9\columnwidth]{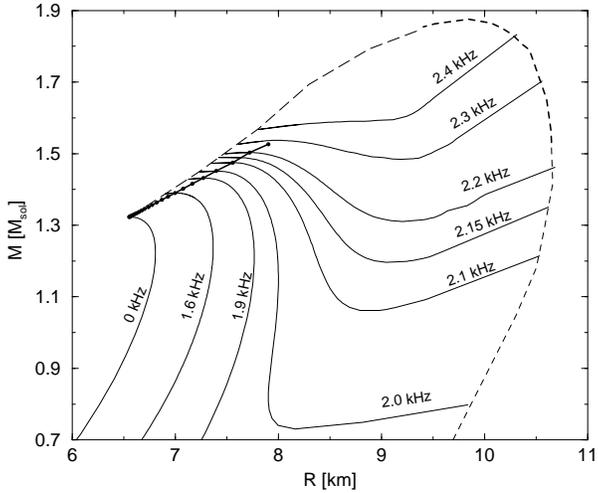}
\caption{Gravitational mass vs. radius for  
  stars described by SS2 model. The thin solid lines correspond to
  sequences of star with constant rotational frequency
  $f=\Omega/2\pi$.  The rotational frequency is labelled close to each
  line. The thick lines (solid and short-dashed) correspond to
  configurations with the maximal mass in each sequence. The dashed
  line corresponds to the mass-shed limit and the long dashed line is
  the quasi-radial stability limit. The intersection of the mass-shed
  and the stability limits give us the locations of the configuration
  rotating with maximum frequency}
\label{MR}
\end{figure}

\begin{figure}
\includegraphics[width=0.9\columnwidth]{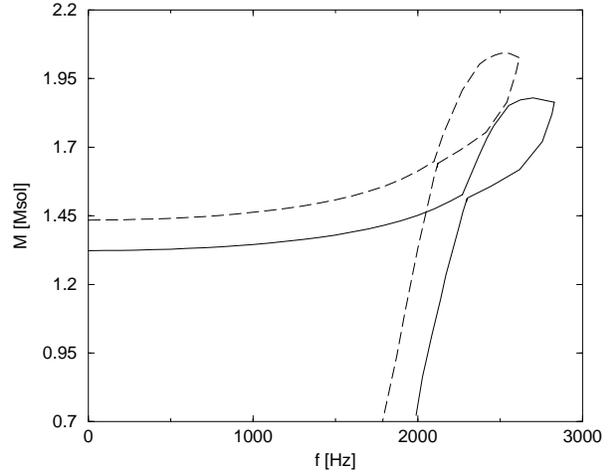}
\caption{Limits on gravitational mass  and the rotation frequency for
  SS1 (dashed lines) and SS2 (solid lines).  
 }
\label{Mfdey1}
\end{figure}

Stable solutions of rotating neutron stars  have to satisfy 
four different constraints \cite{1994ApJ...424..823C}: the static
constraint - for normal evolutionary sequence of rotating stars,
when angular momentum  goes to zero the  configuration should
be  identical to the one described by OV equations for the same
baryon mass;  the low mass constraint below which a neutron star
cannot form, the mass-shed (Keplerian) constraint , and the
constraint of stability to quasi-radial perturbations.

The  first three constraints provide bounds on normal sequences
stars (these are always stable to quasi-radial perturbations),
while the two last limits provide  bounds on the supramassive
stars. 

The mass-shed limit is reached when the velocity at the equator
of a rotating star is equal to the velocity of an orbiting
particle (a star becomes unstable when gravitational attraction
is not sufficient to hold matter bound to the surface). For
normal sequences of neutron stars and low mass supramassive
neutron stars, Keplerian configurations are always these with the
maximal rotational frequency. For high mass supramassive neutron stars
the Keplerian configuration is the one with lowest rotational
frequency in the sequence.

The fourth constraint is the requirement of stability to axisymmetric
perturbations. For an evolutionary sequence parameterized by the central
density, the model is (secularly) stable if $\displaystyle \left( {\partial J
\over \partial \rho_{\rm c}} \right)_{\rm M_{\rm bar}} < 0$ (or $\displaystyle
\left( {\partial M \over \partial \rho_{\rm c}} \right)_{\rm M_{\rm bar}} < 0$)
and unstable otherwise \cite{1986ApJ...304..115F}.  The stability constraint
imposes a limit which begins at the maximum mass static configuration and
terminates on the Keplerian limit sequence near the maximum mass rotating
configuration (see for example gravitational mass versus central density
dependence at Figure 1 in \cite{1994ApJ...424..823C}). 
 Normal sequences begin at the
static limit and terminate at the mass-shed limit. Along such sequences angular
momentum increases while central density decreases. Supramassive star sequences
begin at the stability limit with a minimal angular momentum. As the angular
momentum increases, the central density decreases until the configuration
terminates at the Keplerian limit.  The intersection of the mass-shed and the
stability limits on the gravitational mass - central density plane give us the
locations of the configuration rotating with maximum frequency. For neutron
stars the maximum mass Keplerian configuration is not necessarily stable
against axisymmetric perturbations. For some equation of states of neutron
stars the maximum mass configurations is the same as the configurations with
the maximum frequency.

 \subsection{Limits on gravitational mass}
 
To find equilibrium sequences of compact strange stars we take
only three of the above constraints into account: the low mass
constraint is not relevant for self-bound matter.  Other
instabilities (for example to nonaxisymmetric perturbation) are
not considered here since we cannot study them with our code. 
Taking into account all constraints described above we found
limits on masses and rotation frequencies for the SS2 model,
shown as thick lines in Figure~\ref{Mfdey2}.  The mass shed
limit is shown as a thick short-dashed line, the stability limit
as long-dashed thick line, the fastest normal and low mass
supramassive configurations as a dot-dashed line, and the "low
rotational frequency" maximum mass configurations as a thick
solid line.  The thin solid lines in Figure~\ref{Mfdey2}
correspond to normal and massive supramassive equilibrium
evolutionary sequences and are labelled with their  baryon mass.
In Figure~\ref{Mfdey2cd} we show evolutionary sequence with
baryon masses $1.8 M_\odot$ as an example of the low mass
supramassive stars (with baryon mass lower than 1.9 $M\odot $).
The marginally stable configurations with respect to quasi-radial 
perturbation is marked as open circle. The angular momentum increases 
along each curve from $J=0$ for static configurations ($J_ {\rm min}$ for
supramassive stars) to $J_{\rm max}$ for the Keplerian ones
represented by filled circles for normal and low mass
supramassive sequences.  The evolutionary sequence with $M_{\rm
bar}=1.64$ separates normal stars from the supramassive ones.

The maximum mass configuration is found by considering a sequence of
stars with a constant rotational frequency $f=\Omega/2\pi$ and
parameterized by the central density.  The maximum mass models for
each sequence are represented by thick lines in Figure ~\ref{MR}.
Chosen sequences are shown as thin solid lines.  The Keplerian
configurations are shown as a dashed line, while marginally stable
configurations with respect to quasi- radial oscillation as
long-dashed line.  On one end of each sequence there is a Keplerian
configuration (with the lowest central density in the sequence), and
on the other end the last stable configuration with respect to
axisymmetric perturbations (the densest object in the sequence). For
$f< 2.27\,$kHz the maximum mass configuration (shown as thick solid
line) for each sequence is close to the marginally stable one. For
fast rotating configurations, $f> 2.27\,$kHz the maximum mass
configurations are the Keplerian ones (for comparison see figure 1b in
\cite{2000puas.conf..661G} for strange stars described by the MIT Bag
model).

The thick solid line and thick short dashed line in
Figure~\ref{Mfdey2} correspond to the maximal gravitational mass of
rotating strange star described by the SS2 equation of state as a
function of the rotational frequency. There are two regimes as the
rotational frequency increases, first for rotational frequency lower
than $2.27\,{\rm kHz}$ (in the case of SS2 model) and second for $f >
2.27\,{\rm kHz}$.  In the first regime the maximal mass limit line
goes very close to the stability limit one (shown as thin long dashed
line).  For rotation frequencies lower than $1\,$ kHz the stellar
configuration is only slightly affected by rotation - the increase in
the maximal mass is only a second order effect (where both the maximum
mass line and the normal evolutionary sequence with maximum baryon
mass are very close to each other) ie.  $M_{\rm max}(f) - M_{max}^0
\propto f^2$, where $M_{max}^0$ is the static mass configuration. For
higher frequencies $M_{\rm max}(f)$ increases faster.  The maximum
mass of rotating compact strange star in this regime is approximately
$15\%$ greater than the maximum mass of a static stellar configuration.
For $f>2.27\,$kHz we see a strong increase of the maximum mass (thick
dashed line). In this regime the configurations with maximum mass are
Keplerian.  The maximum mass of a rotating compact strange star in
this regime (the maximum allowed mass of rotating configurations) is
$42\%$ greater than the maximum mass of a static stellar configuration.

Note, that the maximal mass configurations for a given rotational
frequency are always supramassive i.e. no static configuration with
such a mass can exist.  It is worth to note that the central density
of the rotating configurations is always lower than the central
density of the maximum mass static star. This makes us confident that
the rotating configurations found with the use of approximation of
equation~\ref{linea} are at least as accurate as the static
configuration calculations.

\subsection{Limits on rotational frequency}

Let us now consider upper limits on rotational frequency of
compact strange stars. In the case of high mass supramassive
stars the maximum rotational frequency is determined by the
condition of stability to axisymmetric perturbations shown as a
thick long dashed line in Figure \ref{Mfdey2}.  Similarly to
the supramassive neutron stars, the configurations to the left
side of this line are stable to radial collapse - they spin up
loosing their angular momentum as discussed by \cite{1994ApJ...424..823C} 
 in the case of supramassive neutron stars and by
\cite{1999A&A...349..851G} in the case of supramassive MIT Bag
model strange stars.

In the case of normal and low mass supramassive stars, the maximum
rotational frequency is slightly above the rotation frequency of the
Keplerian configuration. The fastest configurations are shown as a
thick dash-dotted line in Figure~\ref{Mfdey2} while the Keplerian one
by filled circles.  In Figure~\ref{Mfdey2cd} we present the details of
the intermediate region of Figure~\ref{Mfdey2} for $2.0\ {\rm kHz} < f
< 2.5\  {\rm kHz}$ to show low mass supramassive stars sequences. 
 
The rotational frequency decreases for large values of the angular
momentum and the sequences "turn back" in Figure~\ref{Mfdey2} and
Figure ~\ref{Mfdey2cd} before reaching the Keplerian configuration. At
this small part of an evolutionary sequence, configurations spin up by
loosing the angular momentum (or slows down obtaining the angular
momentum).  The mass-shed configurations are reached due to the
increase of the equatorial radius related to the deformation of the
rotating star. The difference between the Keplerian frequency and the
maximal rotation frequency for these evolutionary sequences SS2 model
is of the order of $2\%$.  Such a phenomenon was discussed by Zdunik et al. 
(2000) in the case of normal sequences of MIT Bag model strange
stars.  It is interesting to note that no such behavior was noticed
for neutron stars. This feature is characteristic for stars described
by a self-bound EOS.

The mass shed limit and the stability limit lines intersect twice: at
frequency $\approx 2.25\,$kHz and at the frequency of $\approx
2.8\,$kHz.  The intersection of the mass-shed limit (solid thick line
in Figure~\ref{Mfdey2}) and stability limit (solid dashed line in
Figure~\ref{Mfdey2}) determines the configuration rotating with
maximum allowed frequency. Note that the maximum frequency occurs for
only one extreme supramassive model and that this model is at the
mass-shed limit and stability limit.

The configuration with the absolute maximum mass lies on the 
Keplerian limit line. Note that the maximum mass configuration is not
rotating with the maximum allowed rotational frequency.  For
both SS1 an SS2 models the maximum mass  rotating configuration 
is on the stable side of mass-shed limit line.

In Figure~\ref{Mfdey1} we present  the regions of the parameter
space  given by mass and rotation frequency  where strange stars
described by the SS1 (dashed line) and SS2 (solid line)
equations of state can exist. The maximal mass and maximal
frequency configurations lie on the  Keplerian limit line.  
Details of the   Keplerian configurations with the maximum mass
are given in the right columns of Table 1.  The
maximum mass of rotating strange star given by SS1 and SS2 model is by 
$42 \% $ (by $50 \% $ the  equatorial radius) larger than in the 
case of nonrotating star with maximum mass.  For strange stars with
massless and not interacting quarks these values are $44 \% $
and $54 \% $ respectively. The ratios of masses $M_{\rm
max}^{\rm rot}/M_{\rm max}^{\rm stat}$ and equatorial radii
$R_{\rm max}^{\rm rot}/R_{\rm max}^{\rm stat}$ do not depend on
the parameter $\rho_0$, since static and rotating maximum mass
configurations scale identically, and are  functions only of 
$a$ (sound velocity) in the case of stars described by equation 1.

For neutron stars the maximum mass which can be supported when
the star is rotating uniformly increases by $14  \%$ to $21  \%$
depending on the equation of state while the  radius increases
by $30 \%$ to $39 \%$ (e.g. \cite{1994ApJ...424..823C,1998A&A...334..943D}. 
 We note that a large  increase (higher than in
the case of neutron stars) of the maximum mass and corresponding
equatorial radius in the case of strange stars of different EOS
due to rotation is related to the fact that these EOS are
self-bound with a very high density at the surface. They are
much more compact than neutron stars and  much higher rotation
frequencies are required  to deform them to reach Keplerian
configurations.

\begin{figure}
\includegraphics[width=0.9\columnwidth]{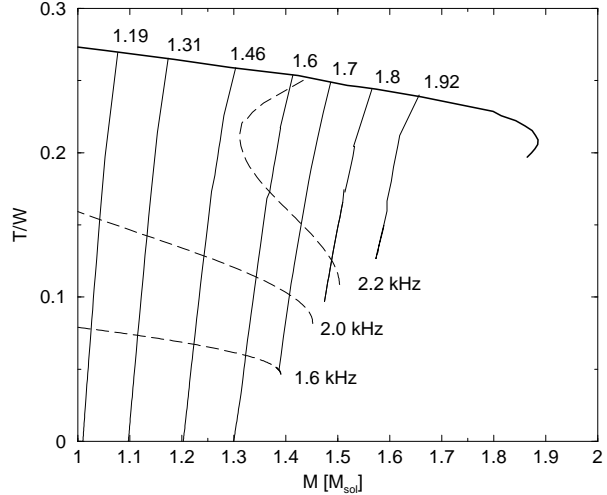}
\caption{The ratio of the rotational kinetic energy to the 
  absolute value of the gravitational potential energy $T/W$ for SS2
  evolutionary sequences with fixed baryon mass labelled above each
  line in sollar mass units. The solid thick line corresponds to
  Keplerian configurations. The thin dashed line correspond to
  sequences of configuration rotating with the same rotational frequency. 
Models located to left and below of this line rotate with lower frequency.
}
\label{ptw}
\end{figure}

\subsection{$T/W$ ratio for rotating strange stars}
It has been shown by Gourgoulhon et al. 1999 and Stergioulas et al.
1999 (see table 1) that $T/W$ can be very high for strange stars
described by the simplest MIT Bag model.  In Figure~\ref{ptw} we
present the ratio of the rotational kinetic energy to the absolute
value of the gravitational potential energy $T/W$ for evolutionary
sequences for SS2 model.  Each sequence is labeled with its baryon
mass. The thin dashed line corresponds to the sequence of
configurations rotating with the same rotational frequency.  Note,
that for a rotating strange star the value of $T/W$ is significantly
higher than that for an ordinary NS (eg. Cook et al. 1994). This
property is universal for all self-bound EOS. $T/W$ does not depend on
$\rho_0$ and its dependence on $a$ is very weak.  We can easily obtain
a similar figure for the simplest MIT model EOS of SS using the
scaling relations $M_{\rm SS}=(a_D \rho_{0,ss}/a_{\rm SS}\rho_{0,\rm D})^{-1/2}
M_{\rm D}$,$f_{\rm SS}=(a_{\rm D} \rho_{0,ss}/a_{\rm SS}\rho_{0,\rm
  D})^{1/2}f_{\rm D}$ (Stergioulas et al 1999, for Keplerian
configurations), where the subscript $D$ denotes the EOS considered
here.  We check that above scaling relations holds for any models in
evolutionary sequences. The difference in mass and rotational frequency 
between numerical
results and rescaled ones are of the order of $2 \%$ and $4\%$. The value of
$T/W$ is quite large for all configurations close to Keplerian ones,
so it is possible that the point of onset of secular instability to
non-axisymmetric normal modes has already been passed.  For mass-shed
configurations $T/W$ varies from 0.21 for maximum mass configuration
to 0.27 for the low mass one (0.25 and 0.26 for gravitational mass 1.4
$M_\odot$ for SS1 and SS2 Keplerian models).  This could be an
indicator that rapidly rotating SS may constitute strong sources of
gravitational waves (Gourgoulhon et al. 1999, Gondek-Rosi\'nska et
al.,2000).

\section{Discussion}

\subsection{Summary of the properties of rotating compact strange
stars}

We have calculated numerical models of the uniformly rotating strange
stars described by the SS1 and SS2 equations of state using the
multi-domain spectral methods, which allows to treat the density
discontinuity at the surface of "self-bound" stars with even very high
$\rho_0$. The model used here is describes the quark interactions
self-consistently. The maximum mass of strange stars within this
model is relatively low, and the stars are  very compact. 
 We find that the stars within the \cite{1998PhLB..438..123D}
  model can rotate
much faster than typical neutron stars, and also the MIT bag
model strange stars. The maximum allowed rotational frequency 
is $2.6\,$kHz for SS1 and $2.8\,$kHz for SS2 .
The maximal mass of a rotating configuration is $1.87\,M_\odot$ for SS1
and for SS2.  The main physical reason for these high value
of the rotational frequency and the low gravitational mass
is that the parameter $\rho_0$ is quite large for this EOS's.

 Some properties of rotating strange star described by the
 \cite{1998PhLB..438..123D}  EOS  are universal and  characteristic for all
self-bound EOS. We find that: i) there are two regimes for
maximal mass of a rotating configuration as the rotation
frequency increases; first, for low rotation frequencies the
increase in the maximal mass is only second order effect
($M_{\rm max}(f)$ is very close to the line of limiting
stability against quasi-radial perturbations limit); second, for
higher frequencies the maximal mass configurations are
Keplerian;  ii), the maximum mass of strange stars given by the
MIT model and by SS1 and SS2 at the point of intersection with
the line imposed by the Keplerian limit the is approximately
$15\%$ greater than the maximum mass of a static static
configuration; iii),  the maximum allowed mass is approximately 
$40 \% $  larger than the static maximum mass. This is much 
higher than  for  neutron stars; iv), we show that in contrast
to normal neutron stars the maximal rotating frequency for both
normal and supramassive stars is never the Keplerian one;  v),
we find that rotating strange stars   have a very high ratio
$T/W$. In the case of the  Keplerian limit stars the ratio 
$T/W$ increases with decreasing  mass. Large values of
$T/W$ (higher than 0.2) imply that it  is quite likely that the 
maximum rotational frequency can in fact be lower than found hear.

\subsection{Astrophysical aspects of the compact strange stars}

The maximum frequency is very high - 2.6 kHz  and 2.8 kHz
for SS1 and SS2 models, respectively. It is important to
remember that the maximum frequency occurs for only one extreme
supramassive model and that this model is both on the mass-shed limit
and the stability limit. But even for normal evolutionary sequences
we reach very high frequencies - higher than 1.8 kHz and 2 kHz
in the case of SS1 and SS2 respectively.  The periods for stars
rotating with maximal frequency can be shorter than half millisecond,
much shorter than the period $P=1.56 \ {\rm ms}$ of the fastest
known  millisecond pulsar PSR 1937 + 21.

The maximal masses for the SS1 and SS2 EOS are consistent with the
observed masses of compact object. All observed pulsars have masses
close to $1.4 \, M_\odot$ and rotate with frequencies lower than the
maximal frequencies for the SS1 and SS2 models.  In the case of
strange stars described by the SS1 and SS2 equations of state the
maximum baryon mass are $1.64$ and $1.85$ (the difference between the
baryon and gravitational mass is much greater than in the case of neutron 
stars).  One can speculate that for high central densities in
the core of a neutron star a phase transition to strange matter can
take place. This can be accompanied by large energy release, and
possibly a gamma-ray burst \cite{1996PhRvL..77.1210C,2000ApJ...530L..69B}.
 A rotating neutron star may become a strange star conserving
its total baryon mass and angular momentum. If the baryon mass of this
star is lower than 1.84 $M_\odot$ it would become a normal sequence
compact strange star.  Otherwise, depending on its angular velocity,
it can become a stable supramassive strange star and after it slows
down finally a black hole.  Just before transformation from a
supramassive strange star to a black hole the star should spin up
(such phenomena was noticed by \cite{1994ApJ...424..823C}
 in the case of
supramassive neutron stars).  In Figure~\ref{Mfdey2cd} we show an
evolutionary sequence with the baryon mass $1.8 M_\odot$ as an example
of a low mass supramassive stars. This sequence begins at high $J$ on
the Keplerian limit, then losing angular momentum it reaches the
maximum mass limit, and finally it spins up to reach the stability
limit and collapse to a black hole.  Assuming that a neutron star goes
through a strange star stage and then ends up as a black hole may be
an explanation why we do not observe pulsars with masses much higher
than 1.4 $M_\odot$ (if pulsars are strange stars described by SS1 and
SS2 equation of state.)

The masses of compact objects in LMXBs inferred from kHz QPOs
and assuming that the highest QPO frequency observed is related
to marginally stable  orbit are quite large, and extend even
above $2\,M_\odot$. Such high mass neutron stars can still
undergo a phase transition to form a supramassive compact
strange star, which  would consequently turn into a black hole. 
Note that the binary might be disrupted during the  transition. 
To find such stars we would be looking for very fast 
millisecond pulsars, either single or in binaries.

\acknowledgements This work has been funded by the KBN grants
2P03D00418, 2P03D01413, and 2P03D00415 and also made use of the NASA
Astrophysics Data System.  SR, MD and JD are grateful to Abdus Salam
ICTP, the IAEA and the UNESCO for hospitality at Trieste, Italy, and
to Dept. of Science and Technology, Govt. of India for a grant no.
SP/S2/K18/96.  The numerical calculations have been performed on
computers purchased thanks to a special grant from the SPM and SDU
departments of CNRS.

\bibliographystyle{aabib99}


\end{document}